# Plasmonic Light Illumination Creates a Channel to Achieve Fast Degradation of Ti3C2Tx Nanosheets

Jiebo Li[#,†], Ruzhan Qin[∥,†], Li Yan[§], Zhen Chi[‡], Zhihao Yu[∥], Mingjun Hu[§,*], Hailong Chen[‡,*], Guangcun Shan[∥,*],

[#] Beijing Advanced Innovation Center for Biomedical Engineering, School of Biological Science and Medical Engineering, Beihang University, Beijing, 100083, China

[∥] School of Instrumentation Science and Optoelectronics Engineering, Beihang University, Beijing, 100191, China

[§] School of Materials Science and Engineering, Beihang University, Beijing, 100191, China

[‡] Beijing National Laboratory for Condensed Matter Physics, CAS Key Laboratory of Soft Matter Physics, Institute of Physics, Chinese Academy of Sciences, Beijing, 100190, China

[∥] College of Chemistry and Molecular Engineering, Beijing National Laboratory for Molecular Sciences, Peking University, Beijing 100871, China

[†] Equally work;

[*] Corresponding author;

**ABSTRACT:**
Two-dimensional (2D) material-controllable degradation under light radiation is crucial for their photonics and medical-related applications, which are yet to be investigated. In this paper, we first report the laser illumination method to regulate the degradation rate of $Ti_3C_2T_x$ nanosheets in aqueous solution. Comprehensive characterization of intermediates and final products confirmed that plasmonic laser promoting the oxidation was strikingly different from heating the aqueous solution homogeneously. Laser illumination would nearly 10 times accelerate the degradation of $Ti_3C_2T_x$ nanosheets in initial stage and create many smaller-sized oxidized products in a short time. Laser-induced fast degradation was principally ascribed to surface plasmonic resonance effect of $Ti_3C_2T_x$ nanosheets. The degradation ability of such illumination could be controlled either by tuning the excitation wavelength or changing the excitation power. Furthermore, the laser- or thermal-induced degradation could be retarded by surface protection of $Ti_3C_2T_x$ nanosheets. Our results suggest that plasmonic electron excitation of $Ti_3C_2T_x$ nanosheets could build a new reaction channel and lead to the fast oxidation of nanosheets in aqueous solution, potentially enabling a series of water-based applications.

## Introduction

Since this decade, MXene has been a new promising series of 2D materials with excellent conductivity, hydrophilicity and mechanical properties.[1-3] As structure view[4], MXenes as the general formula $M_{n+1}X_nT_x$ are composed of stacked 2D sheets where M stands for an early transition-metal carbide (Ti, Ta, Mo, Nb, V, and Zr), X is either carbon or nitrogen, $T_x$ represents surface functional groups such as O, OH, and/or

F. Thus, MXene with ultra-thin atomic layer thickness exhibits abundant physical and chemical properties. And, the precise and controllable preparation of MXene elements and structural units provides a more extensive and flexible material science foundation for the multifunctional exploration of MXene. Therefore, based on the structure, Mxene could have many exciting applications in energy storage and conversion[5,6], water splitting[7], water desalination[8] and steaming[9], electronics,[10] electrostatic shielding[11,12], cancer therapy[13], and thermal imaging[14,15].

To utilize Mxene for beneficial usage, versatile physical properties are employed of MXene, such as photothermal conversion. With outstanding energy conversion efficiency, light-to-heat conversion of Mxene has gained renewed research interest in the recent researches and found itself in energy and medical applications.[16-23] A recent work reported that MXene showed promising photothermal properties in converting light energy for water steam as practical solar energy utilization.[9] Liu et al. and Lin et al. both investigated MXene's application to phototherapy under near-infrared and their results show outstanding in vitro/in vivo photothermal ablation performance of MXene on tumor cells.[21,22] XX also reported Mxene could also be used as photoacoustic imaging agent for guiding therapy, showing Mxene could be promising theranostic agents in future.[15] However, very recently, Lotfi et al predicted the oxidation stability of MXene structures in wet air.[24] And Zhang et al that the MXene could be degrading slowly in colloidal solution.[25] Thus, the stability of MXene colloidal solutions, under light and heat condition, become a critical important question to their energy conversion and medical applications to be explored.

In this work, we report the investigation of MXene stability in varies conditions under changing temperature, light irradiation, synthesis materials and surface coating. These processes are monitored by UV−vis, SEM, XPS and HR-TEM. We found that both heat and light conditions could significantly promote the degradation of MXene in aqueous solution. The proposed mechanism of degradation reaction is based on the thermal driven with the correlation between the degradation kinetics and the temperature. Structure with more defects are more easily to degrade. Finally, we demonstrated that well surface layer coating could strongly slow down the degradation under heating condition.

## Results and Discussion

Characterization of $Ti_3C_2T_x$ Nanosheets and Degradation Products under Laser Illumination. In this work, the lasers with different wavelengths and powers were employed as light sources for the investigation of light stability of typical $Ti_3C_2T_x$ aqueous dispersion solution. As a contrast, the degradation of $Ti_3C_2T_x$ solution under heating was also studied by employing a separate heat source, such as an oven.

Figure 1a shows a schematic diagram of laser irradiation through $Ti_3C_2T_x$ solution, where the laser power remained constant in the entire illumination process and the solution temperature was measured by an infrared thermometer. The structures and compositions of $Ti_3C_2T_x$ nanosheets before and after laser irradiation and heating were investigated. XRD patterns clearly showed the changes of crystalline structure of the powders in different stages. Distinguished from raw materials of MAX phase, the strong diffraction peak at ∼6° was presented in the exfoliated sample; this is much less than 9.5°, which corresponds to the diffraction angle of (002) face of MAX ceramics, indicating effective exfoliation of $Ti_3AlC_2$ powders. Figures 1b and 1c showed SEM images of the exfoliated products, and well-defined multilayer structure and a typical few-layer 2D structure were presented, denoting that high-quality $Ti_3C_2T_x$ nanosheets were formed. Figure S2 in the Supporting Information displayed that many delaminated $Ti_3C_2T_x$ nanosheets were distributed on silicon substrates, and these nanosheets had an average size of 2.5 ± 1 μm in the lateral direction. For fresh delaminated $Ti_3C_2T_x$ nanosheets (see Figures 1d and 1e), an interplanar spacing of ∼5.35 Å could be identified in the HRTEM image, which was much larger than the interspace of the (004) face in the corresponding MAX phase with lattice distance of 4.62 Å (JCPDS No. 00-052-0875). The results revealed that Al atom layers have been well-removed and the bond of MXene nanosheets has been dominated by van der Waals force.[43,44]

After 5 h of laser irradiation (808 nm, 0.7 W), the structure of $Ti_3C_2T_x$ nanosheets changed significantly, with many $TiO_2$ nanoparticles appearing on the surface of nanosheets (Figure 1f). The crystalline nanostructures were exhibited in Figure 1g, and these nanoparticles showed the interplanar spacing of 2.05 Å, which corresponded to the (210) crystal face in rutile $TiO_2$, but not matching any crystal face in anatase phase, indicating that the main product derived from laser-induced oxidation was rutile $TiO_2$, which was also supported by our XRD and Raman results (see Figures S1a and S1b in the Supporting Information). In the Raman spectra of the laser-illuminated sample, the characteristic peaks located at 435.9 and 623.7 cm$^{-1}$ could be assigned to rutile $TiO_2$.[33,45] Further oxidation led to nearly complete conversion of low-valence titanium into $TiO_2$, and the solution finally turned cloudy white. Figures 1h and 1i showed TEM and HRTEM images of the degradation products of $Ti_3C_2T_x$ nanosheets after laser illumination, followed by long-time aging, where $TiO_2$ nanoparticles became bigger and the number further increased, and most of them were located on a layer of 2D membrane, which was suggested to be amorphous carbon.[29] Scanning electron microscopy (SEM) was adopted to analyze the morphology evolution of $Ti_3C_2T_x$ nanosheets before and after heating and laser illumination. Figure 2 presented the morphology of fresh $Ti_3C_2T_x$ nanosheets. After 5 h heating at 40 °C, small amount of oxide nanoparticles appeared at the edges and cracks of nanosheets (Figure 2b), meaning that oxidation happened first at defect sites. In contrast to the heated sample, significantly more oxide nanoparticles could be observed in laser-illuminated samples, which is indicative of a rapider oxidation rate under laser illumination. In addition, we can also see that the nanosheets after laser illumination seem to present a smaller size (Figure 2c) than that observed after 5 h of heating and may be attributed to laser-induced generation of microcracks on nanosheets. After the following room-temperature standing of a few days, the products would turn cloudy white, which indicates the generation of many $TiO_2$ nanoparticles, which was also verified by SEM imaging (Figure 2d).

X-ray photoelectron spectroscopy were used to characterize the chemical composition of original $Ti_3C_2T_x$ nanosheets and oxidative products. XPS survey spectra of $Ti_3C_2T_x$ samples with different oxidation states were presented in Figure 2e, and the presence of several main elements, including Ti, C, O, N, and F, was confirmed. After different oxidation durations, the intensity of O 1s peak was enhanced gradually and F 1s peak was weakened, indicating that Ti−F bonds were broken and new Ti−O bonds were formed with the increase of titanium valence states. High-resolution XPS spectra of $Ti_3C_2T_x$ in Ti 2p region and C 1s region were performed to study the change of the valence and circumstance of titanium and carbon elements during oxidation. Ti 2p core-level XPS spectrum of original $Ti_3C_2T_x$ nanosheets was displayed in Figure 2f, and the spectrum was fitted into four pairs of peaks (Ti $2p_{3/2}$ and Ti $2p_{1/2}$) with a fixed area ratio of 2:1. Four Ti $2p_{3/2}$ peaks were located at ∼455.2, 456.3, 457.5, and 459.7 eV respectively, with the area percentages of 39%, 33%, 20%, and 8%, respectively. According to some previous reports,[29,35,46,47] these peaks can be assigned to Ti−C, Ti(II) suboxides and/or hydroxides, Ti(III) suboxides and/or hydroxides, and

Ti(IV) oxides, respectively, indicative of dominant Ti−C component in original $Ti_3C_2T_x$ nanosheets and little oxidation. After 5 h of laser illumination, we can see that the subpeaks of Ti−C and Ti(II) components decreased, and Ti(III) and Ti(IV) peaks were enhanced significantly (Figure 2g), indicating that rapid conversion of titanium from low valence to high valence states under laser irradiation. In contrast to a heat-treated sample (Figure 2h), the Ti(II) component in the laser-illuminated sample experienced a visibly quicker oxidation process into Ti(III) and Ti(IV) oxides while the degradation rate of Ti−C components did not display prominent difference, which is consistent with the XPS fitting result from carbon 1s core-level spectrum (Figure S3 in the SI).

Accordingly, Ti(IV) oxides and Ti(III) suboxide components in the Ti 2p XPS spectrum of the 5-h laser-illuminated sample comprised a relatively larger percentage than that of the heated sample, indicating that laser is a more efficient mode to induce the conversion from Ti(II) suboxides to high-valence titanium.

However, when the duration was extended to 1 week, laser-illuminated $Ti_3C_2T_x$ nanosheets did not degrade more rapidly than that by heat irradiation, and even in reaction residues, the laser-illuminated sample has more low-valence titanium residues (see Figures 2i and 2j). It may be due to slight aggregation of the $Ti_3C_2T_x$ nanosheets after illumination oxidation, which results in poorer irradiation efficiency and the loss of plasmonic features of oxidized products that led to the weakening of the reactivity. In addition, we also fitted carbon 1s core-level XPS spectra of different $Ti_3C_2T_x$ oxidative products with six subpeaks according to some previous reports,[46] corresponding to C−Ti (281.9 eV), C−Ti−$T_x$ (282.8 eV), C=C (284.0 eV), C−C (285.1 eV), C−O comparing the C 1s XPS spectra between a laser-irradiated sample and a heat-irradiated sample (see Figures S3b and S3c), we suggested that, with the oxidation of low-valence titanium, oxygen molecules ($O_2$) captured electrons from surface titanium atom and weakened the C−Ti bonds. With the continuous attack of oxygen, C−Ti bonds would break completely, generating $TiO_2$, and a new π bond would form to produce C☐C bonds, which was the reason why the amount of C☐C ($sp_2$) bonds increased markedly and C−C bonds decreased. Figures S3b and S3c also show that the 5-h laser-irradiated sample possessed a bigger share of C−C component than the heat-irradiated sample, probably because of the formation of the C−H bond derived from the reaction of C☐C bonds with hydrogen or hydrogen free radicals under high-power laser irradiation. With the continuous oxidation of $Ti_3C_2T_x$ nanosheets, the laser effect dropped off, and the oxidation rate decreased obviously, leaving a small amount of unoxidized residues in the laserilluminated sample.

As control studies, XPS spectra of the products of the resulting $Ti_3C_2T_x$ solution treated by 5 h of laser illumination, followed by 2 weeks of room temperature aging, as well as 5 h of heating at 40 °C plus room temperature aging for 2 weeks were also investigated. It was found that the degradation of $Ti_3C_2T_x$ nanosheets in these cases was more complete, in contrast to the sample illuminated for 1 week by laser, and the peak that was assigned to the Ti−C component almost disappeared in both samples, but in relative terms, the amounts of Ti−C residues in the laser-illuminated sample were still greater than those observed for the heated one. The residues were mainly composed of amorphous carbon, $TiO_2$, and a small amount of titanium carbides and Ti(II) suboxides. The survival of low valence titanium might be due to the protection of amorphous carbon membrane due to strain-induced conformal package, which, however, would not happen under heat irradiation, because of a slow reaction rate and timely stress release. In addition, according to high-resolution C 1s core-level spectra in these two samples, obviously more C=C components could be found in the laser-illuminated sample than in the heated one, in accord with the results observed for the sample that was laser-illuminated for 1 week, which was indicative of more obvious oxidation of C−C bonds under laser illumination, which was also consistent with Raman results in Figures S1d and S1e.

Degradation Dynamics of $Ti_3C_2T_x$ Nanosheets under Laser Illumination and Heating. To quantitatively understand the laser-induced degradation phenomenon, UV-vis-NIR spectra measurements of the solution were performed. As shown in Figure 3a, $Ti_3C_2T_x$ nanosheets in water present three distinctive peaks. With increasing laser illumination time, the intensity of the 260 nm peak would increase, and the intensities of the 325 and 770 nm peaks would decrease during degradation. According to the Lambert−Beer law, the $C_{MXene}$ is proportional to the peak intensity. To capture the degradation rates comparable with references,[29] the intensity at the plasmonic peak center was chosen to identify the concentration of $Ti_3C_2T_x$ nanosheets. Thus, it was suitable for choosing the peak intensity at 770 nm, as a function of time, to track the degradation of the colloidal solution.

图 1 RT, heat laser

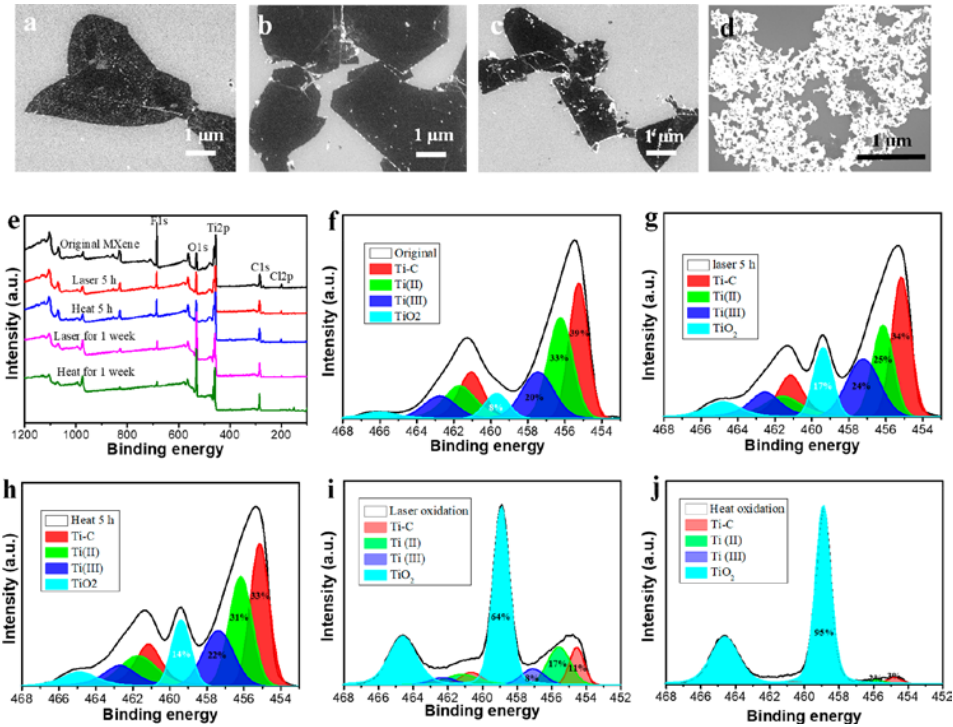

Figure 2. SEM images of Ti$_3$C$_2$T$_x$ samples: (a) initial sample, (b) after heating for 5 h, (c) after laser illumination for 5 h, and (d) after aging for 2 weeks. (e) XPS survey spectra of Ti$_3$C$_2$T$_x$ samples under different oxidized states. XPS spectra of Ti 2p core level in Ti$_3$C$_2$T$_x$ samples with different oxidized states: (f) original Ti$_3$C$_2$T$_x$ nanosheets, (g) after laser illumination for 5 h, (h) after heating at 40 °C for 5 h, (i) after laser illumination for 1 week, and (j) after heating for 1 week at 40 °C.

| Sample | Time Constant (hours) |
|---|---|
| 40 °C continues | 311 |
| 50 °C continues | 217 |
| 60 °C continues | 96 |
| 800 nm Laser Irradiation then RT | 38 |
| 40 °C heating then RT | 273 |

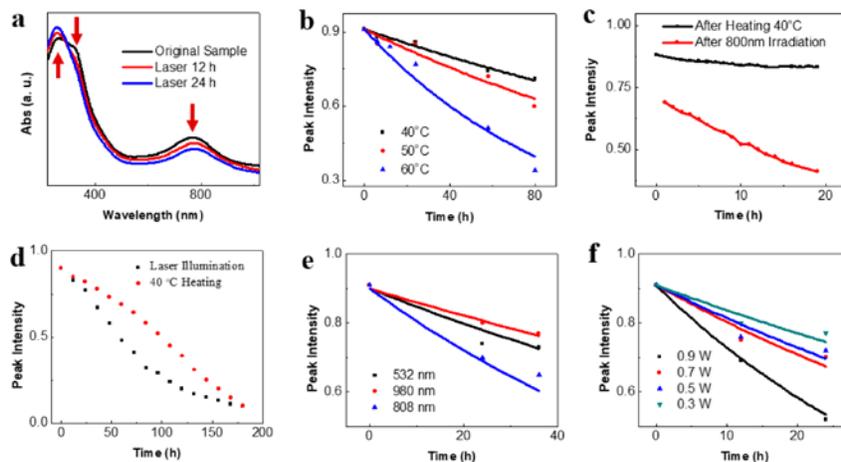

Figure 3. UV-vis-NIR spectra of Ti$_3$C$_2$T$_x$ in solution under different conditions: (a) full spectrum after laser illumination; (b) stability results under heating at 40, 50, and 60 °C; (c) stability results after femtosecond laser illumination; (d) stability results under 808 nm CW laser illumination and heating at 40 °C; (e) stability results under 532, 980, and 808 nm CW laser

illumination; and (f) dependence on laser intensity.

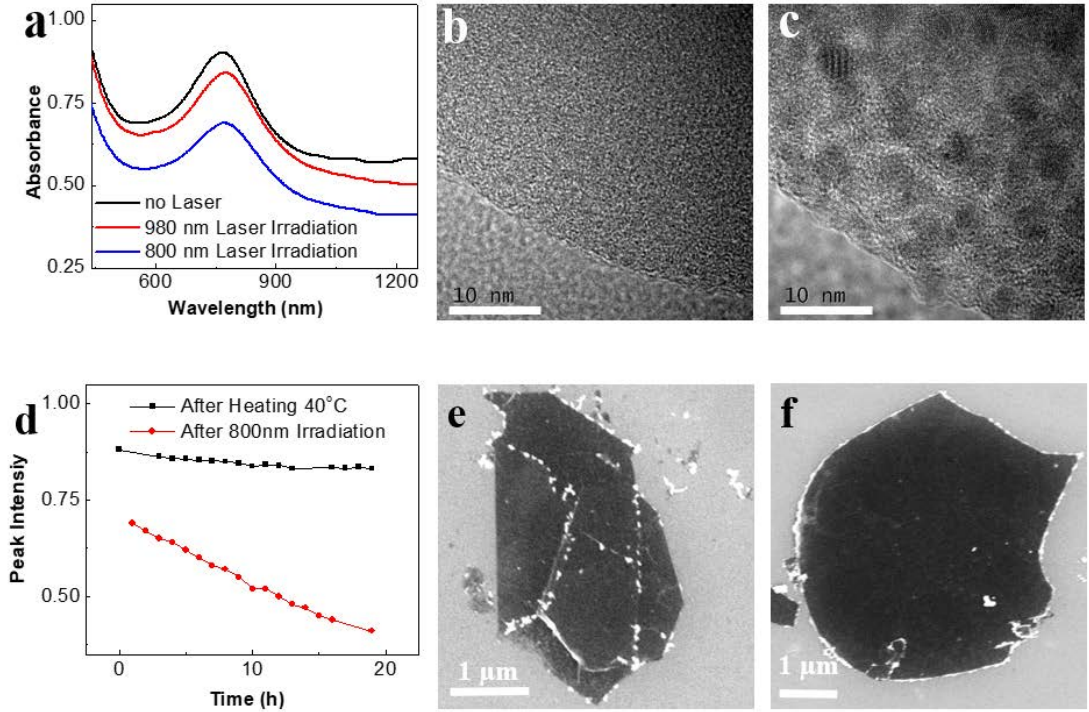

图 4 机理示意图；

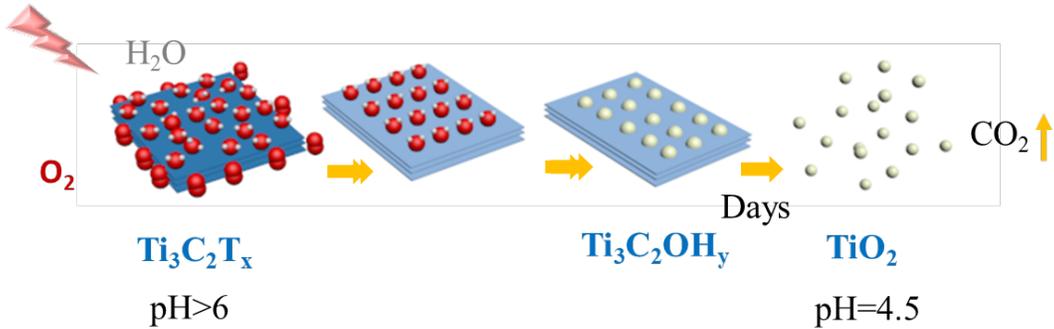

图 5 结构导致速率差别

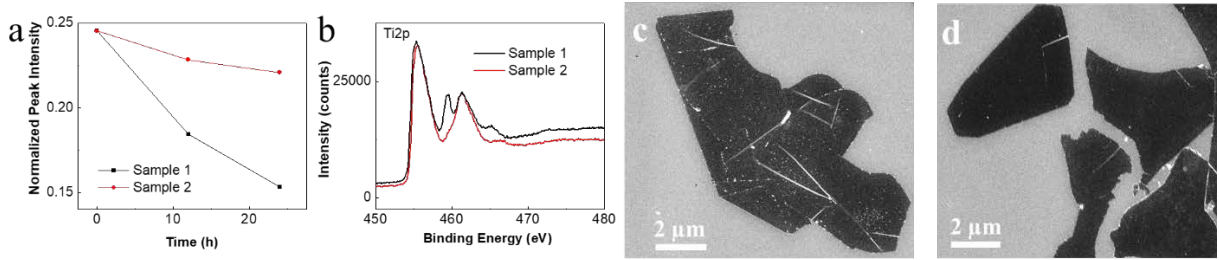

## 图 6 表面保护显著减慢降解

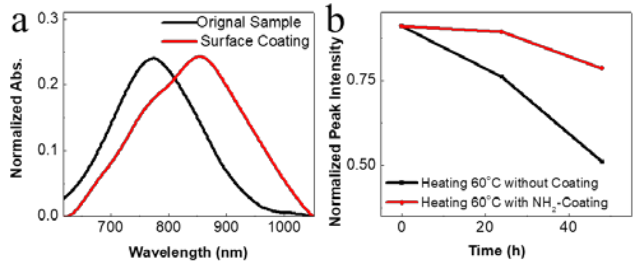

1. Light and heat accelerate the dissociation

    1.1 describe Light and heat promote the reaction

    1.2 temperature dependent reaction
    1.3 light frequency selection

2. Difficult structure has distinctive dissociation rate
   In this work, to understand the dissociation mechanism, we evaluated two kinds of mxene structures for heat induced dissociation. As shown in fig. 4, the two sets of mxene samples are from two sets of materials. These two samples have different visible spectrum. As shown in fig. 4a, sample 1 showed the absorption peak at 760 nm and sample 2 showed the absorption peak at 780 nm.

   Then, we compared the dissociation rates for the two samples. As shown in figure 5, sample 1 peak intensities decreased to 74% after 12 hours and to 62% after 24 hours. In contrast, sample 2 still have 90% after 24 hours. The XPS and SEM showed that the two samples have different structures. Sample1…Sample 2…

   Different storage time sample also changed

3. Surface OH protection decrease the dissociation rate

Shifting the absorption peak. And slow down the degradation rate.

# Plasmonic Light Illumination Creates A Channel to Achieve Fast Degradation of $Ti_3C_2T_x$ nanosheets

Jiebo Li[#,£,†], Ruzhan Qin[∥,†], Li Yan[§], Zhen Chi[‡], Zhihao Yu[□], Naitao Li[#], Mingjun Hu[§,*], Hailong Chen[‡,*], Guangcun Shan[∥,*]

[#] Beijing Advanced Innovation Center for Biomedical Engineering, Beihang University, Beijing, 100191, China

[£] School of Biological Science and Medical Engineering, Beihang University, Beijing, 100191, China

[∥] School of Instrumentation Science and Opto-electronics Engineering, Beihang University, Beijing, 100191, China

[§] School of Materials Science and Engineering, Beihang University, Beijing, 100191, China

[‡] Beijing National Laboratory for Condensed Matter Physics, CAS Key Laboratory of Soft Matter Physics, Institute of Physics, Chinese Academy of Sciences, Beijing, 100190, China

[□] College of Chemistry and Molecular Engineering, Beijing National Laboratory for Molecular Sciences, Peking University, Beijing, 100871, China

\* Corresponding author;

mingjunhu@buaa.edu.cn; gcshan@buaa.edu.cn; hlchen@iphys.ac.cn

**Including figure S1-S12.**

**Interface temperature roughly estimation;**

**Fitting parameters;**

**Roughly estimation of initial MXene ($Ti_3C_2Ti_x$) flake's temperature under CW laser illumination**

Following the paper published by Roper et al [1] and Lin et al [2], the total energy balance for the system was

$$\sum_i m_i C_{p,i} \frac{dT}{dt} = Q_{MXene} + Q_{dis} - Q_{surr} \quad (1)$$

If we consider the temperature of the surface layer water could represent the surface temperature of MXene flakes, we could rewrite the equation (1) as following:

$$dT = (Q_{MXene} + Q_{dis} - Q_{surr}) \times dt / (m_{water} \times C_{water}) \quad (2)$$

Here, $Q_{MXene}$ represented water layer energy transferred from electron-phonon relaxation of MXene flakes surface under laser excitation. $Q_{dis}$ was defined as the baseline energy inputted by the sample cell, and $Q_{surr}$ was sample surface heat dissipation. The electron-phonon process occurred in ultrafast time scale. In our research[3], we measured that >80% energy of MXene could transfer to water at interface in 4 ps. Thus, we roughly assumed that $Q_{dis}$ and $Q_{surr}$ were negligible.

We also assumed in this time scale, surface layer water molecules are defined as the water molecules that directly interact with MXene. Thus, we roughly assumed that one surface site exposed could interact with one water molecule. After ultrasonic MXene in aqueous solution, MXene molecules were well dispersed. The thickness could be 2 nm which means it could be two MXene layers in aqueous solutions. Here, we could roughly calculate the water mass which were binding with MXene surface as following:

$$m_{MXene} / \text{weight}(Ti_3C_2T_x) = m_{water} / \text{weight}(H_2O) \quad (3)$$

Here, weight ($Ti_3C_2T_x$) could be 200-204 (3Ti+2C+2(F or OH or O)). And, weight ($H_2O$) should be 18. Also, $m_{MXene}$ respected the MXene in solution which were directly absorb light. Laser beam had 3 mm diameter and the bottle was 10 mm diameter. The concentration of the solution was 0.05 mg/mL. Thus, $m_{MXene}$ could be calculated as following:

$m_{MXene} = 0.05$ mg/mL $\times 0.07$ mL $= 0.0035$ mg; $m_{water} = m_{MXene} \times 18/202 = 3.14 \times 10^{-7}$ g.

Based on Lin's et al paper, the total energy from laser alse could be expressed as following:

$$Q_{MXene} = I(1 - 10^{-A_{808}})\eta \quad (4)$$

Here, I was incident laser power (700 mW), $A_{808}$ was the absorbance (0.9) of the MXene nanosheets at wavelength, and η was the photothermal conversion efficiency 0.85. Substituting all these values to equation (4), $Q_{MXene}$ was obtained as 0.5 W.

$dT = (Q_{MXene}) \times dt / (m_{water} \times C_{water}) = 0.5$ J/s $\times 4$ ps $/ (3.14 \times 10^{-7}$ g $\times 4.2$ J/(g×°C)) $= 0.15 \times 10^{-5}$ °C;

Thus, the initially, MXene surface temperature could not be elevated by CW laser. The heat could be accumulated after 4 ps, thus, the surface layer could then transfer energy out of interface.

**Roughly estimation of initial MXene ($Ti_3C_2Ti_x$) solution temperature under CW laser illumination**

For our setup, the laser illuminated only small part of the vial, the heating effect thus was small. And monitoring the degradation rate, we detected the whole vial to take spectrums. We could also use the given values η (30.6%) by the pioneers[2] to calculate the initial 1 s, the temperature in the aqueous solutions. Here, we still roughly ignored $Q_{dis}$ and $Q_{surr}$ to calculate the maximum heat accumulation. For the setup in figure S9, the total water volume was 6.3 mL (calculated by 3.14×1/4×8).

Thus, we could obtain dT=0.2 J/s × 1 s / (6.3 g × 4.2 J/(g×°C) = 0.007 °C.

In above calculation, we have two uncertain values: photothermal conversion efficiency and $Q_{surr}$. Even if we selected 10 mins (600 s) as initial time without took consideration of heat emission from sample cell ($Q_{surr}$), the solution temperature only raised up 4.2 °C. It is the reason our thermometer could not detect the temperature raise up significantly for CW laser illumination. Thus, by compared with heat effect, 40 °C was enough for control experiments to heat up the whole vial. For Lin *et al*'s paper[2], even if they employed lower laser intensity, they measured temperature in 96-well culture plate. So, they could obtain higher solution temperature.

Therefore, by compared with the interface temperature roughly estimated by our experiment, we could explain that CW laser illumination promoted the reaction rates better than heating the homogenous sample. The average photothermal conversion might not directly contribute to the reaction degradation.

**Roughly estimation of initial MXene ($Ti_3C_2Ti_x$) flake's temperature under femtosecond laser illumination**

If the incident laser was femtosecond laser, we could consider the experimental condition shown in figure S7. The incident laser intensity was high enough. The single pulse energy was $0.9 \times 10^{-3}$ J (calculated by 0.9 W/1000 Hz). The pulse duration was 120 fs. In total, there were $0.6 \times 10^{-3}$ J energy per pulse converting into thermal energy. From the setup, the value $m_{MXene}$ was calculated as $1.6 \times 10^{-6}$ g.

Then, we could obtain dT in initial pulse from equation (2) as

dT = $(0.6 \times 10^{-3}$ J$)/ (1.6 \times 10^{-6}$ g $\times 4.2$ J/(g×°C)) = 89 °C.

Thus, the initially, MXene surface could be elevated to really high temperature by femtosecond laser. In experiment, we measured the temperature of entire vial around 40 °C.

For femtosecond laser illumination, based on the above calculation, the single pulse could create dT to 89 °C. The starting temperature was 20 °C. Thus, the surface temperature of MXene flakes could reach to 109 °C. Substituting this value into Arrhenius equation, the initial reaction time at surface was obtained as 3 hours. In our experiment, femtosecond laser illumination 5 hours could bring the total absorption from 0.90 to 0.67. Thus, the apparent reaction rate constant was 16 hours. Compared with 280 hours degradation constant under 40°C, the femtosecond intense laser illumination could over 10 times speed up the reaction. Thus, we could not separately rule out either photothermal or hot electron mechanisms for femtosecond laser illumination. Therefore, both mechanisms may contribute to the MXene degradation under 800 nm femtosecond laser illumination.

**Fitting Parameters**

Because our laser experiments are performed in different setups, within one sub-figure, the results are comparable. All fittings in this manuscript are using single exponential decay.

The fitting parameters for figure 3e:

808 nm: time constant 90 h; 980 nm: time constant 220 h; 532 nm: time constant 168 h.

The fitting parameters for figure 3f:

0.9 W: time constant 45 h; 0.7 W: time constant 80 h;

0.5 W: time constant 90 h; 0.3 W: time constant 120 h.

The fitting parameter for adding surfactant under laser illumination (808 nm, 0.7 W): time constant 380 h.

The fitting parameter for coating $NH_2$ under laser illumination (808 nm, 0.7 W): time constant 800 h.

The fitting parameter for coating $NH_2$ under heating 60 °C : time constant 420 h

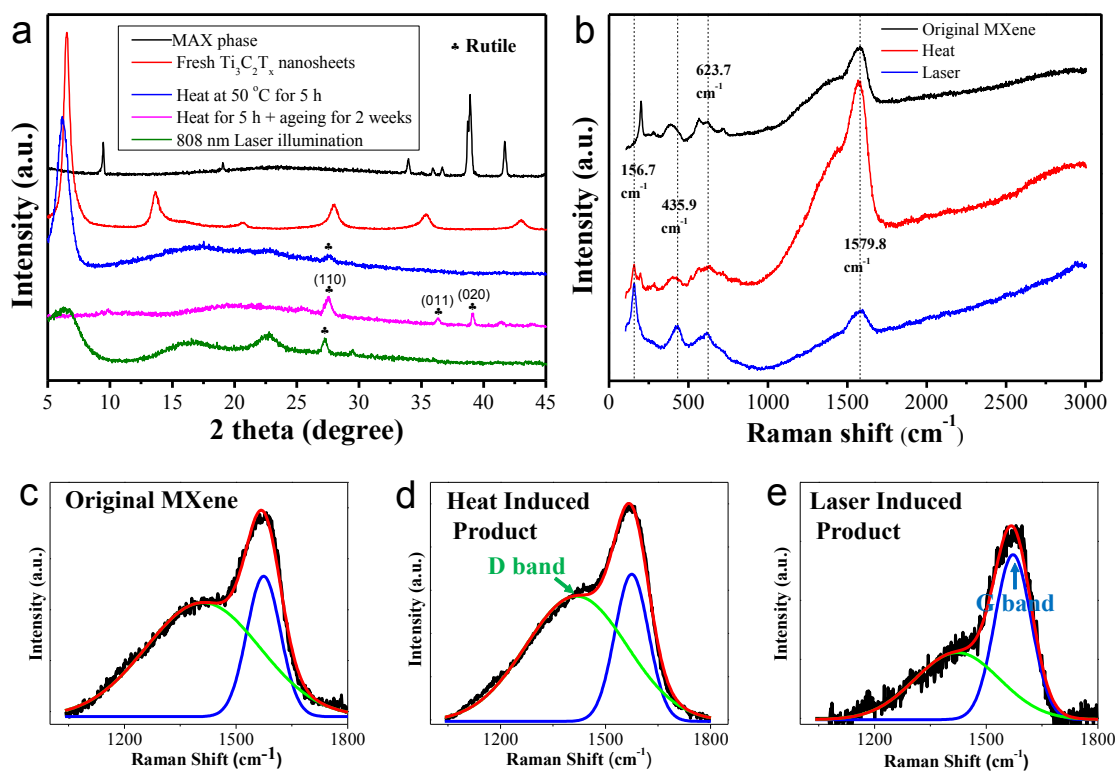

**Figure S1** a) XRD patterns of $Ti_3AlC_2$ MAX powders, delaminated MXene ($Ti_3C_2Ti_x$) nanosheets and degraded products; b) Raman spectra of MXene ($Ti_3C_2Ti_x$) nanosheets and the fully degraded products under heat and laser illumination; c, d, e) Raman spectra of carbon in MXene ($Ti_3C_2Ti_x$) nanosheets and the fully degraded products under heat and laser illumination. For XRD patterns, heating temperature was set to 50 °C, and after heating the sample was sealed in vial to age for 2 weeks.

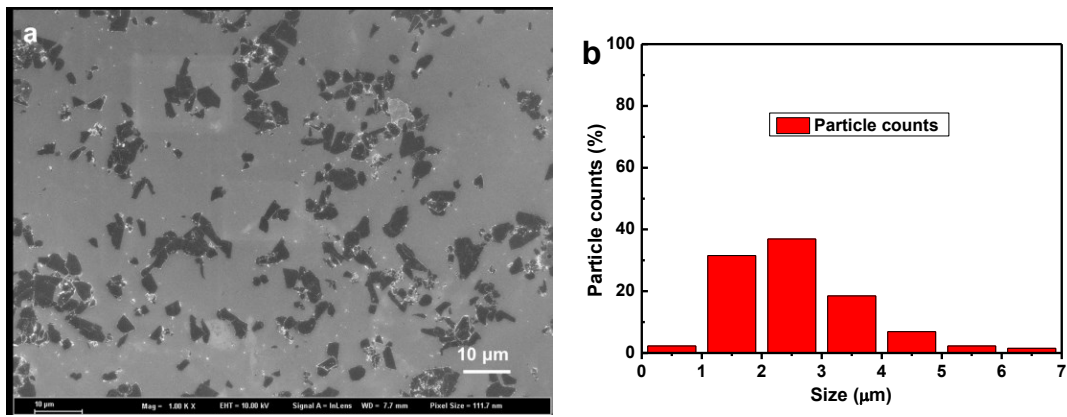

**Figure S2** a) SEM image to describe the size distribution of MXene (Ti$_3$C$_2$Ti$_x$); b) histogram statistic of particle size distribution.

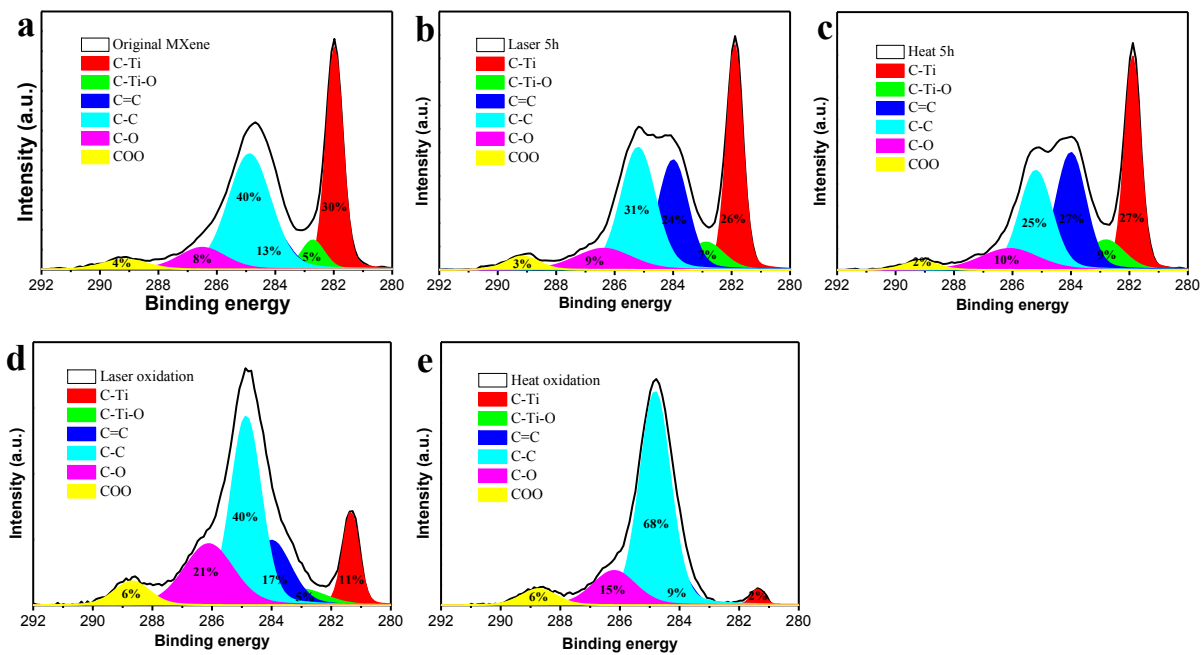

**Figure S3** XPS spectra of MXene (Ti$_3$C$_2$Ti$_x$) in Carbon 1s region with different oxidation states, a) original MXene (Ti$_3$C$_2$Ti$_x$) nanosheets; b) MXene (Ti$_3$C$_2$Ti$_x$) nanosheets after 800 nm laser illumination of 5 h; c) MXene (Ti$_3$C$_2$Ti$_x$) after heating of 5 h at 40 °C; d) MXenes after 800 nm laser illumination for one week; e) MXene (Ti$_3$C$_2$Ti$_x$) after 40 °C heating for one week.

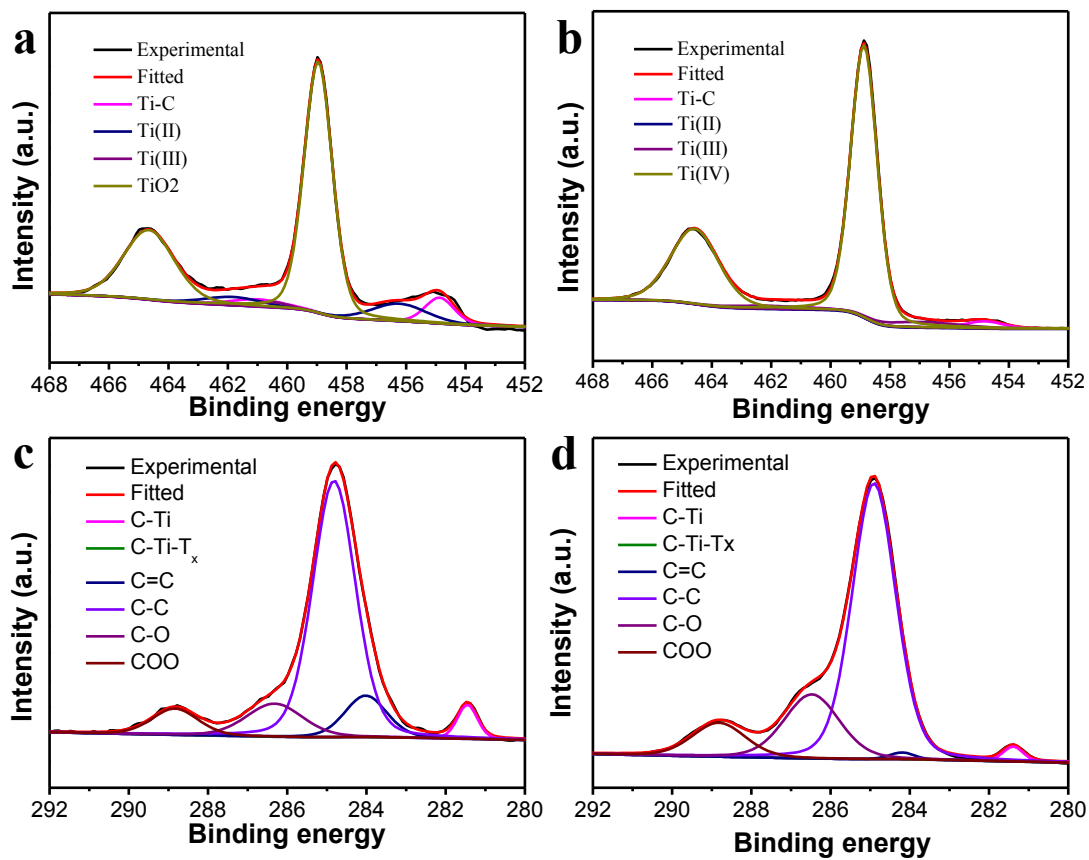

**Figure S4** XPS spectra of MXene ($Ti_3C_2Ti_x$) oxidative products in Ti 2p region and carbon 1s region under light and heat irradiation, a) Ti 2p core-level XPS spectra after laser illumination for 5 h and then ageing for two weeks; b) Ti 2p area after heating at 40 °C for 5 h and ageing for two weeks; c) Carbon 1s area after laser illumination for 5 h and ageing for two weeks; d) Carbon 1s area after heating for 5 h and ageing for two weeks.

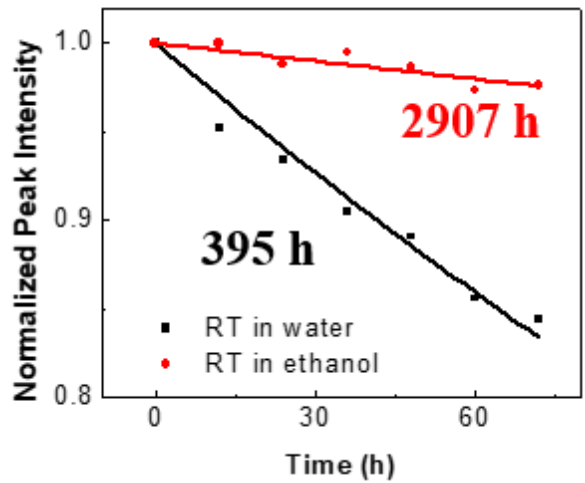

**Figure S5** Kinetic data of MXene (Ti$_3$C$_2$Ti$_x$) in water and ethanol at room temperature. Dots are experimental data and lines are fitting curves.

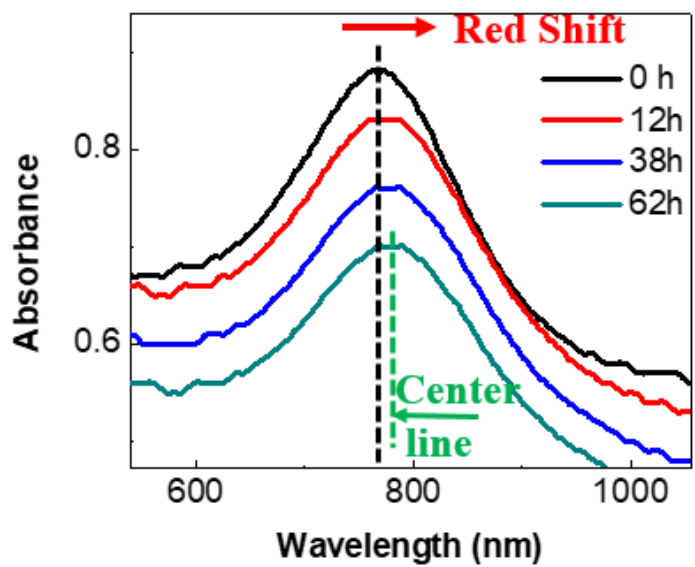

**Figure S6** Vis-NIR spectra of MXene (Ti$_3$C$_2$Ti$_x$) in different heating (at 40 °C) duration.

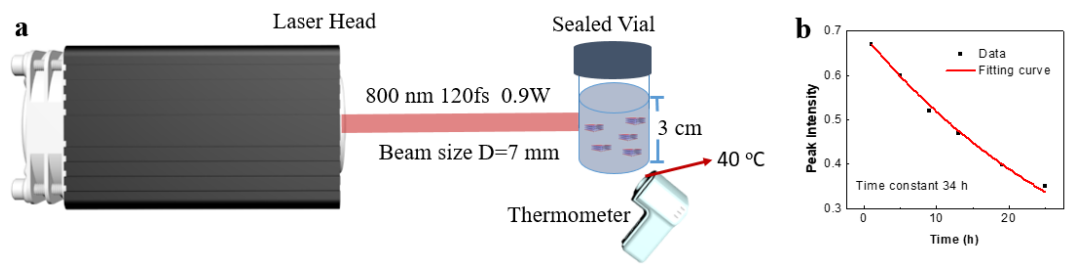

**Figure S7** a) Schematics of femtosecond laser irradiation; b) Kinetic data of MXene ($Ti_3C_2Ti_x$) after 800 nm laser illumination. Before laser illumination, the absorption peak intensity was 0.9. Dots are experimental data and line is the fitting curve.

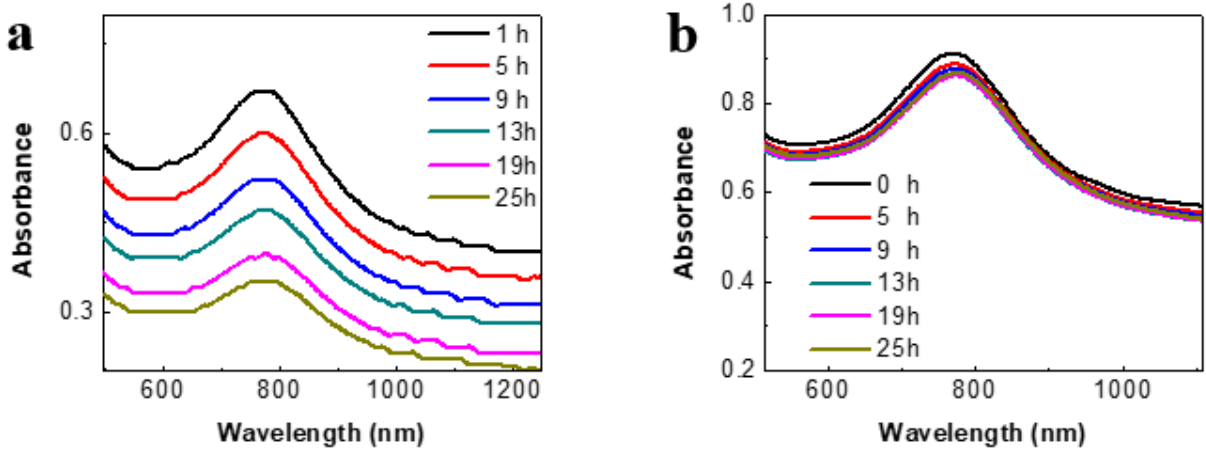

**Figure S8** a) Vis-NIR spectra of MXene ($Ti_3C_2Ti_x$) at RT after 800 nm laser illumination for 5 h and then ageing for different time; b) Vis-NIR spectra of MXene at RT after 40 °C heating for 5 h and ageing for different time.

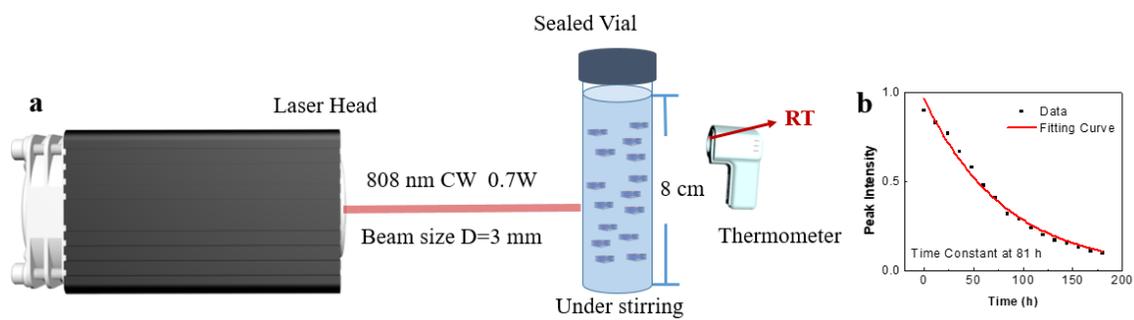

**Figure S9** a) Schematics of 808 nm CW laser irradiation. b) Kinetic data of MXene ($Ti_3C_2Ti_x$) during 808 nm laser illumination. Dots are experimental data and line is the fitting curve.

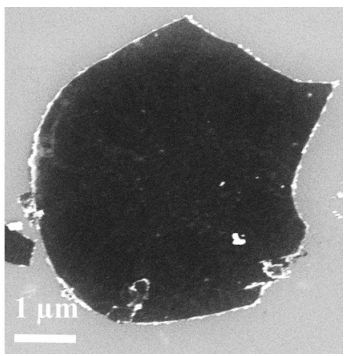

**Figure S10** SEM images of MXene ($Ti_3C_2Ti_x$) nanosheets after 980 nm laser illumination for 5 h.

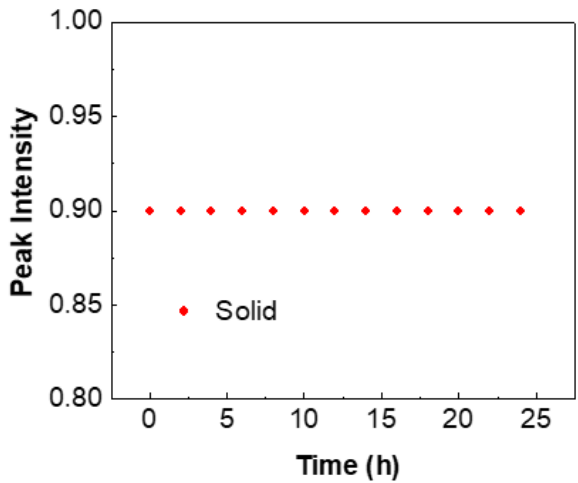

**Figure S11** Kinetic data of MXene (Ti$_3$C$_2$Ti$_x$) as solid transparent film at room temperature.

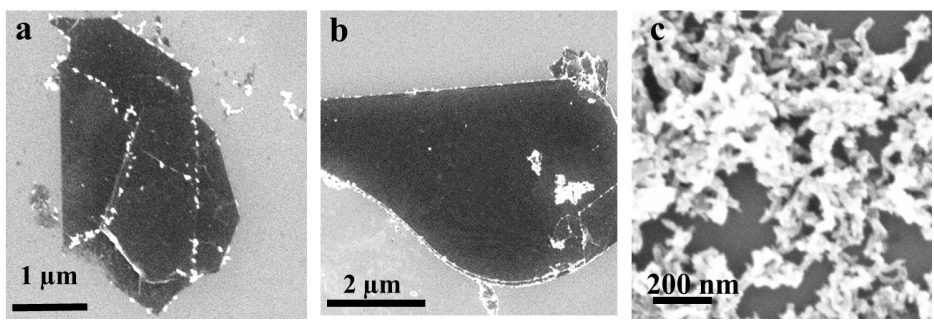

**Figure S12** SEM images of MXene (Ti$_3$C$_2$Ti$_x$), a) after femtosecond laser illumination for 5h; b) after heating at 50 °C for 5 hours; c) after aging for 2 weeks.